\documentclass[sn-mathphys]{sn-jnl}

\jyear{2021}%

\theoremstyle{thmstyleone}%
\theoremstyle{thmstyletwo}%

\theoremstyle{thmstylethree}%

\usepackage{graphicx}
\usepackage{epsfig}
\usepackage{rotate}
\usepackage{amsmath}
\usepackage{amssymb}
\usepackage{amsfonts}
\usepackage{bm}
\usepackage{tablefootnote}
\usepackage{enumerate}
\usepackage{afterpage}
\usepackage{xcolor}
\usepackage{natbib, hyperref}

\newcommand{\red}[1]{{\color{red}{#1}}}

\begin{document}

\title[S. Jolicoeur et al.]{Constraining primordial non-Gaussianity by combining next-generation galaxy and 21 cm intensity mapping surveys}


\author*[1]{\fnm{Sheean} \sur{Jolicoeur}}\email{jolicoeursheean@gmail.com}

\author[1,2,3]{\fnm{Roy} \sur{Maartens}}\email{roy.maartens@gmail.com}

\author[1]{\fnm{Simthembile} \sur{Dlamini}}\email{simther4111@gmail.com}

\affil*[1]{\orgdiv{Physics \& Astronomy}, \orgname{University of the Western Cape}, \orgaddress{\state{Cape Town}, \postcode{7535}, \country{South Africa}}}

\affil[2]{\orgdiv{Institute of Cosmology \& Gravitation}, \orgname{University of Portsmouth}, \orgaddress{\state{Portsmouth}, \postcode{PO1 3FX}, \country{United Kingdom}}}

\affil[3]{\orgdiv{National Institute for Theoretical \& Computational Sciences (NITheCS)}, \orgaddress{\state{Cape Town}, \postcode{7535}, \country{South Africa}}}


\abstract{Surveys of the matter distribution contain `fossil' information on possible  non-Gaussianity that is generated in the primordial Universe. This primordial signal survives only on the largest scales where cosmic variance is strongest. By combining different surveys in a multi-tracer approach, we can suppress the cosmic variance and signficantly improve the precision on the level of primordial non-Gaussianity.
We consider a combination of an  optical galaxy survey, like the recently initiated DESI survey, together with a new and very different type of survey, a 21\,cm intensity mapping survey, like the upcoming SKAO survey. 
A Fisher forecast of the precision on the local primordial non-Gaussianity parameter $f_{\mathrm{NL}}$, shows that this multi-tracer combination, together with  non-overlap single-tracer  information, can deliver precision comparable to that from the CMB. Taking account  of the largest systematic, i.e. foreground contamination in intensity mapping, we find that
$\sigma(f_{\mathrm{NL}}) \sim 4$.}

\maketitle

\section{Introduction}
Constraining the local primordial non-Gaussianity (PNG) parameter $f_{\mathrm{NL}}$ 
presents an important challenge for next-generation large-scale structure surveys.  
Local-type PNG affects the  power spectrum of  dark matter tracers by inducing a scale-dependent correction to the tracer bias \citep{Matarrese:2008nc,Dalal:2007cu}. This correction is strongly suppressed on small to medium scales, but on very large scales, it is $\propto f_{\mathrm{NL}}(H_0^{2}/k^{2})$. This means that  ultra-large survey volumes are required for high-precision constraints. 

The Planck survey of the cosmic microwave background (CMB) gives the current state-of-the-art constraint $\sigma(f_{\mathrm{NL}}) =5.1$ from the bispectrum \citep{Akrami:2019izv}. 
Future CMB surveys will lead to an improvement on the Planck precision, but not by much and not sufficient to access $\sigma(f_{\mathrm{NL}}) \lesssim 1$. 
This level of precision would enable us to rule out many single-field and multi-field inflationary scenarios \citep{Alvarez:2014vva,dePutter:2016trg,Meerburg:2019qqi}.
The hope is that this goal can be achieved by using the extra modes in 3-dimensional surveys of the large-scale structure (see e.g. \citep{Giannantonio:2011ya,Camera:2013kpa,Font-Ribera:2013rwa,Camera:2014bwa,Alonso:2015uua,Raccanelli:2015vla}). However, extracting $f_{\mathrm{NL}}$ from these surveys faces formidable difficulties:
\begin{itemize}
    \item 
Observational systematics on very large scales \cite{Leistedt:2014zqa,Fonseca:2020lmi,Cunnington:2020wdu,Rezaie:2021voi,Liu:2020izx,Spinelli:2021emp,Riquelme:2022amo}.
For our simplified Fisher analysis, we neglect these systematics, except for the foreground contamination of  21\,cm intensity mapping (see below).
\item
Astrophysical complications entailed in the modelling of tracer clustering, in particular,  the uncertainties in modelling the scale-dependent bias \cite{Barreira:2022sey,Lazeyras:2022koc}. We use a simplified model since a full treatment requires simulations, beyond the scope of this paper.
\item
A theoretical systematic arising from the neglect of lensing magnification and other relativistic lightcone effects on the power spectrum, which can bias  the best-fit $f_{\mathrm{NL}}$
\cite{Namikawa:2011yr,Bruni:2011ta,Jeong:2011as,Camera:2014sba,Kehagias:2015tda,Lorenz:2017iez,Contreras:2019bxy,Jelic-Cizmek:2020pkh,Bernal:2020pwq,Wang:2020ibf, Maartens:2020jzf,Castorina:2021xzs,Martinelli:2021ahc,Viljoen:2021ocx}. Although the bias on best-fit can be large, the error $\sigma(f_{\mathrm{NL}})$ is not significantly affected and we therefore omit these lightcone effects.
\item
Cosmic (or sample) variance that dominates ultra-large scale measurements. We deal with this problem via a multi-tracer approach -- which also helps to mitigate uncorrelated systematics.
\end{itemize}

Recent constraints on $f_{\mathrm{NL}}$ have used the BOSS galaxy \cite{Cabass:2022ymb} and eBOSS quasar \citep{Castorina:2019wmr} samples, with the tightest constraint to date of $\sigma(f_{\mathrm{NL}}) = 21$ from eBOSS \citep{Mueller:2021tqa}. The much larger volumes of upcoming surveys should facilitate significant improvement on this precision.
But if we rely on individual surveys using the power spectrum of single tracers, it turns out that  $\sigma(f_{\mathrm{NL}}) \lesssim 1$ is not achievable -- even if we neglect systematics and complexities in scale-dependent bias \cite{Alonso:2015uua}. The fundamental problem is cosmic variance. 

A  way to evade cosmic variance is the multi-tracer technique \citep{Seljak:2008xr,McDonald:2008sh,Bernstein:2011ju,Hamaus:2011dq,Abramo:2013awa,Abramo:2015iga,Witzemann:2018cdx,Wang:2020dtd,Abramo:2022qir,Alonso:2015sfa,Fonseca:2015laa}. Forecasts indicate that multi-tracing next-generation surveys can surpass the CMB precision, and in some cases can achieve $\sigma(f_{\mathrm{NL}}) \lesssim 1$ \citep{Ferramacho:2014pua,Yamauchi:2014ioa,dePutter:2014lna,Alonso:2015sfa,Fonseca:2015laa,Fonseca:2016xvi,Schmittfull:2017ffw,Munchmeyer:2018eey,Fonseca:2018hsu,SKA:2018ckk,Gomes:2019ejy,Ballardini:2019wxj,Bermejo-Climent:2021jxf,Viljoen:2021ypp}. The performance of the multi-tracer  improves when the difference in tracer properties, especially the Gaussian clustering biases, is significant. It also helps to suppress uncorrelated systematics. 

In this paper, our aim is not to identify survey combinations that can achieve $\sigma(f_{\mathrm{NL}}) \lesssim 1$ -- for this purpose, one typically requires the very high number densities of photometric surveys \cite{dePutter:2014lna,Alonso:2015sfa}. Instead, our aim is to investigate the improvements over single-tracer constraints when using a new type of spectroscopic large-scale structure survey -- neutral hydrogen (HI) intensity mapping of the 21\,cm emission line -- in combination with spectroscopic galaxy surveys. Such a pair of spectroscopic samples has very different clustering biases and systematics, which accentuates the advantages of a multi-tracer analysis.  

We use a simple Fisher forecast on mock surveys, which are based on next-generation surveys that are starting to observe or close to starting. The two surveys we have in mind are: the  Dark Energy Spectroscopic Instrument (DESI), with the Bright Galaxy Sample (BGS) and Emission Line Galaxy (ELG) samples \citep{DESI:2016fyo,Yahia-Cherif:2020knp}, and  the Square Kilometre Array Observatory (SKAO), with intensity mapping surveys in Bands 1 and 2 \citep{SKA:2018ckk,Berti:2022ilk}. Galaxy number count surveys are well established, but a cosmological HI intensity mapping survey has not yet been implemented. However, pilot surveys on the SKAO precursor telescope MeerKAT have already been used to:\\ (a) measure the cross-power between MeerKAT and WiggleZ galaxy surveys 
\cite{Cunnington:2022uzo}, and\\ (b) start developing a pipeline for the planned SKAO surveys \cite{Santos:2017qgq,Wang:2020lkn}. 

We combine pairs of these surveys at low and high redshifts, using the  Fisher single-tracer constraints in the non-overlap volume and the multi-tracer constraints in the overlap volume. The results show a reasonable improvement for the low-redshift pair, and a significant improvement for the high-redshift pair, compared to the standard single-tracer forecasts for $\sigma(f_{\mathrm{NL}})$. The  combination of all low- and high-redshift surveys  improve on Planck, with $\sigma(f_{\mathrm{NL}})\sim 4$. This constraint is based on avoiding foreground contamination of HI intensity mapping on very large scales. Foreground contamination of HI intensity mapping has a strong effect on the precision:  if we neglect this foreground contamination, we find that $\sigma(f_{\mathrm{NL}})\sim 1$.


\section{Multi-tracer power spectra}
At first order in perturbations,
the density (or temperature) contrast of a tracer  $A$ is related to the matter density contrast $\delta$  by
\begin{equation}
\Delta_A(z, \bm{k}) = \Big[b_A(z)+f(z)\mu^{2}\Big]\delta(z, \bm{k})\,,  \label{e1.7}
\end{equation}
where $b_A$ is the (Gaussian) clustering bias, $\mu = \bm{\hat{k}} \cdot \bm{{n}}$ is the projection along the line-of-sight direction $\bm n$, and $f$ is the linear growth rate.
We assume $b_A$  has a known redshift evolution $\beta_A$, following  \cite{Agarwal:2020lov}, so that
\begin{equation}
b_{A}(z) = b_{A0}\,\beta_A(z)\,,\label{bias1}
\end{equation}
where the amplitude $b_{A0}$ is a free parameter. We highlight the point that  a multi-tracer approach reduces the impact of this assumption on $\sigma(f_{\mathrm{NL}})$ \cite{Viljoen:2021ocx}.

The Fourier power spectra at tree-level are given by
\begin{equation}
\big \langle \Delta_A(z,\bm{k})\Delta_{B}(z,\bm{k}') \big \rangle = (2\pi)^{3}P_{AB}(z,\bm{k})\delta^{\mathrm{D}}(\bm{k}+\bm{k}')\,, \label{e1.10}
\end{equation}
where the linear matter power spectrum (computed using CLASS \citep{Blas:2011rf}) is
\begin{equation}\label{pab}
P_{AB}=\big(b_A+f\mu^2 \big)\big(b_B+f\mu^2 \big)P.    
\end{equation}
In the presence of local PNG, the  bias acquires a scale-dependent correction \red{\cite{Matarrese:2008nc, Dalal:2007cu, Lazeyras:2022koc}}:
\begin{equation}
\hat b_A(z,k) = b_{A}(z) + b_{A\phi}(z)\frac{f_{\rm NL}}{\mathcal{M}(z,k)}\,, \quad \mathcal{M}(z,k) = \frac{2}{3\Omega_{m0}H_{0}^{2}}\frac{D(z)}{g_{\rm in}}T(k)\,k^{2}\,. \label{e1.17}
\end{equation}
Here $T(k)$ is the matter transfer function (normalized to 1 on very large scales), $D$ is the growth factor (normalized to 1 today) and $g_{\rm in}$ is the initial growth suppression function, defined deep in the matter era. For $\Lambda$CDM \cite{Villa:2015ppa} 
\begin{equation}
g_{\rm in} = \frac{3}{5}\big(1+z\big)D(z)\bigg[1+\frac{2f(z)}{3\Omega_{m}(z)}
\bigg]\,. \label{e1.18}
\end{equation}
The non-Gaussian bias factor $b_{A\phi}$ is halo-dependent and should be determined with the aid of simulations \citep{Barreira:2020kvh,Barreira:2021dpt,Barreira:2022sey}. A  simplified  model reduces it to 
\begin{equation}
b_{A\phi}(z) = 2 \delta_{\rm c}\big[b_{A}(z)-p_A\big]\,,\label{e1.19}
\end{equation}
where  the critical collapse density is given by $\delta_{\rm c} =1.686\,$ and $p_A$ are halo-dependent constants. In the simplest (universal) halo model, $p_A=1$ for all tracers $A$. But this model is not consistent with simulations \citep{Barreira:2020kvh,Barreira:2021dpt,Barreira:2022sey}. An improved (but still over-simplified) model allows the constant $p_A$ to vary with tracer.
For galaxies chosen by stellar mass, simulations indicate that the rough approximation \citep{Barreira:2020kvh}
\begin{equation}
p_{g} \approx 0.55\,,\label{eQQ2}
\end{equation}
is an improvement, and we use this for the DESI-like samples.
For HI intensity mapping, we use the approximation \citep{Barreira:2021dpt}
\begin{equation}
p_{H} \approx 1.25\,.\label{eQQ3}
\end{equation}

The $f_{\mathrm{NL}}$ terms in the power spectra are of order $H_0^2/k^2$ on  ultra-large scales, $k\lesssim k_{\rm eq}$, where $T\approx 1$. Local PNG therefore dominates the power on ultra-large scales -- but these are also the scales where cosmic variance is largest.  
We deal with the cosmic variance via a multi-tracer analysis, which includes the information from all auto- and cross-power spectra of the two (or more) tracers (see \autoref{secfish}).

\section{Modelling the surveys}

We consider two mock spectroscopic surveys:

 $A=g$:~ galaxy survey, similar to DESI surveys \cite{DESI:2016fyo,Yahia-Cherif:2020knp};
 
 $A=H$: 21\,cm HI intensity mapping (IM) survey, similar to SKAO surveys \cite{SKA:2018ckk,Viljoen:2020efi,Berti:2022ilk}.

In both cases, we have a low- and a high-redshift survey. The sky and redshift coverage of the individual and overlapping surveys are given in
\autoref{tabInfo}.
\begin{table}[!ht] 
\centering 
\caption{Sky area and redshift range of surveys.} \label{tabInfo} 
\vspace*{0.2cm}
\begin{tabular}{|l|l|c|c|c|} 
\hline 
Survey~~&~~Sample~~  & ~~$\Omega_{\mathrm{sky}}$~~ &~~$t_{\mathrm{tot}}$~~ & ~~redshift~~ \\ 
		&         & $\big[10^{3}\,\mathrm{deg}^{2}\big]$ & $\big[10^{3}\,\mathrm{hr}\big]$ & ~~range~~ \\ 
\hline\hline 
$g$~(DESI-like) & BGS & 14 & - & 0.00--0.50 \\
		    & ELG  & 14 & - & 0.60--1.70 \\ \hline \hline
$H$~(SKAO-like)   & 
Band 2 & 20 & 10 & 0.10--0.58
 \\
		    & 
Band 1 & 20 & 10 & 0.35--3.05
\\ \hline \hline
           $g\times H$~(low $z$) & BGS $\times$ Band 2 & 10 & 5 & 0.10--0.50 \\ 
           $g \times H$~(high $z$) & ELG  $\times$ Band 1 &  10 & 5 & 0.60--1.70 \\ \hline
\end{tabular}
\end{table}

\begin{figure}[!ht]
\centering
\includegraphics[width=8cm]{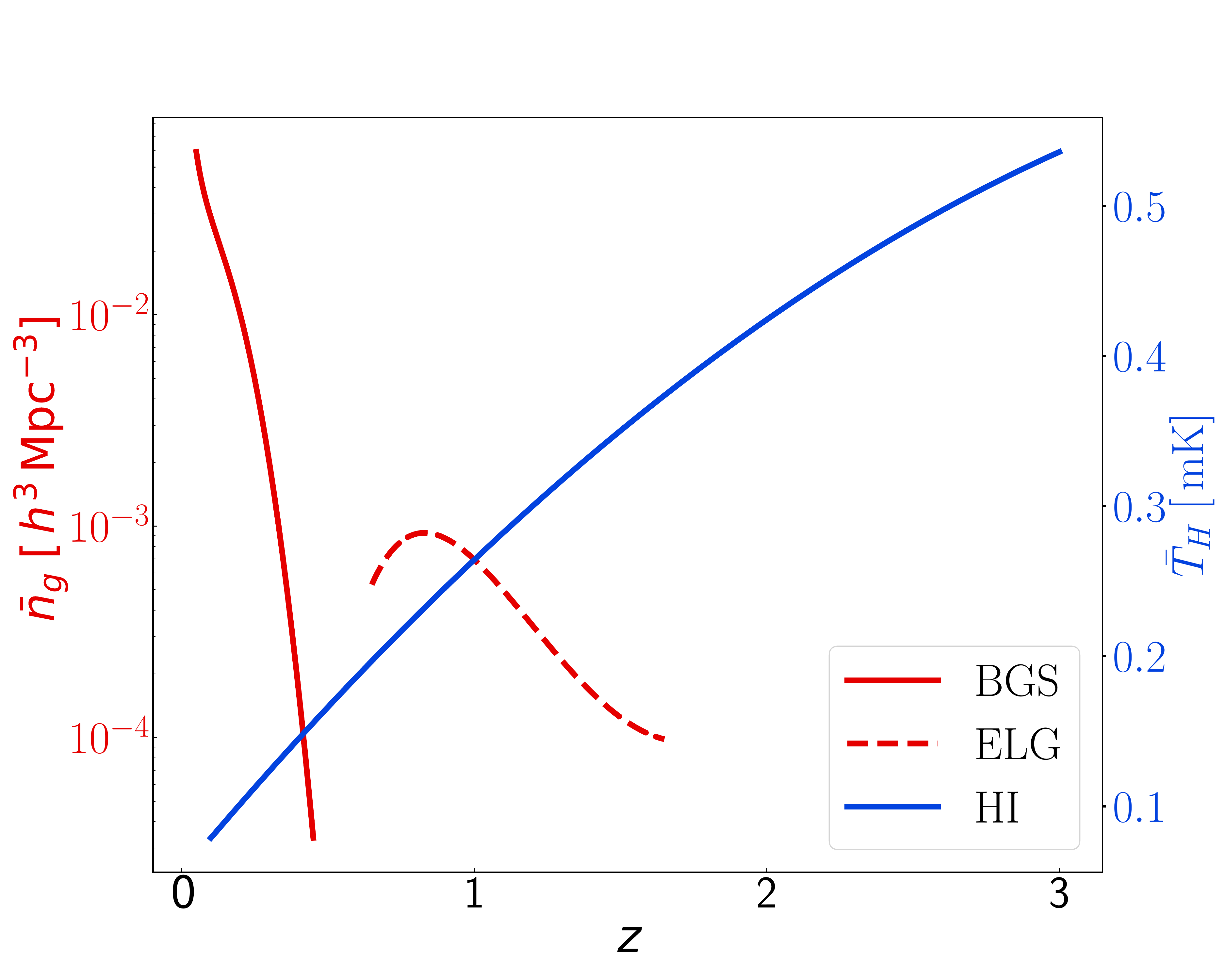} 
\caption{Background comoving galaxy number densities (red, left-hand $y$-axis) and HI IM temperature (blue, right-hand $y$-axis).}\label{fign}
\end{figure}

The one-parameter model \eqref{bias1} for the Gaussian clustering bias of the galaxy surveys follows \citep{Jelic-Cizmek:2020pkh}
\begin{equation}
b_{g}(z) = \frac{b_{g0}}{D(z)} \,. \label{up2}
\end{equation}
Fiducial values are $b_{g0}=1.34$ for the  Bright Galaxy Sample (BGS) and $b_{g0}=0.84$ for the Emission Line Galaxy (ELG) sample. For the HI IM surveys we use a fit based on \citep{Villaescusa-Navarro:2018vsg,Cunnington:2022ryj}:
\begin{equation}
b_{H}(z) = {b_{H0}} {\big(1 + 0.823\,z - 0.0546\,z^{2}\big)}\quad\mbox{with fiducial value}~b_{H0}=0.842 \,.\label{up2}
\end{equation}
The background comoving number density $\bar n_g$ for the BGS and ELG samples is modelled following \citep{Jelic-Cizmek:2020pkh}, which leads to the smoothed curves shown in \autoref{fign}.
For HI IM, the background brightness temperature is modelled via the fit given in \citep{Santos:2017qgq,Fonseca:2019qek}:
\begin{equation}
\bar{T}_{H}(z) = 0.0559 +0.2324\,z -0.0241\, z^{2} ~~ \mathrm{mK}\,, \label{E1.34}
\end{equation} 
which is also shown in  \autoref{fign}.

\subsection{Noise}
In galaxy surveys, the shot noise (assumed to be Poissonian) is
\begin{equation}
P^{\rm shot}_{gg}(z) = \frac{1}{\bar n_{ g}(z)}\,. \label{E1.31}
\end{equation}
Then the total signal that we measure for the galaxy auto-power spectrum is
\begin{equation}
\tilde{P}_{gg}(z,k,\mu) = P_{gg}(z,k,\mu) + P^{\rm shot}_{gg}(z)\,,\label{EE1.32}
\end{equation}
where $P_{gg}$ is given by \eqref{pab}. 

In IM surveys, there is shot noise, but also  thermal (or instrumental) noise. The Poissonian shot noise (see \cite{Umeh:2021xqm} for non-Poissonian corrections) is derived in a halo-model framework and given by \cite{Castorina:2016bfm, Villaescusa-Navarro:2018vsg},
\begin{equation}
P_{H\!H}^{\rm shot}(z)=\frac{1}{\bar n_{H}(z)}
= \frac{1}{\bar \rho_H^{\,2}(z)}\int \mathrm{d}  M\,{\mathcal{N}}_h(M,z)\,M_H^2(M,z)\,, \label{phnois}
\end{equation}
where $\bar n_{H}$ is the {effective average comoving number density} of HI, $\bar \rho_H$ is the average comoving HI density, ${\cal N}_h$ is the halo mass function (average comoving halo number density per mass) and $M_H$ is the average HI mass in a halo of mass $M$. 
However, on the linear scales that we consider, the  shot noise is much smaller than the thermal noise and can be safely neglected \cite{Castorina:2016bfm, Villaescusa-Navarro:2018vsg}.

\begin{figure}[! ht]
\centering
\includegraphics[width=5.9cm]{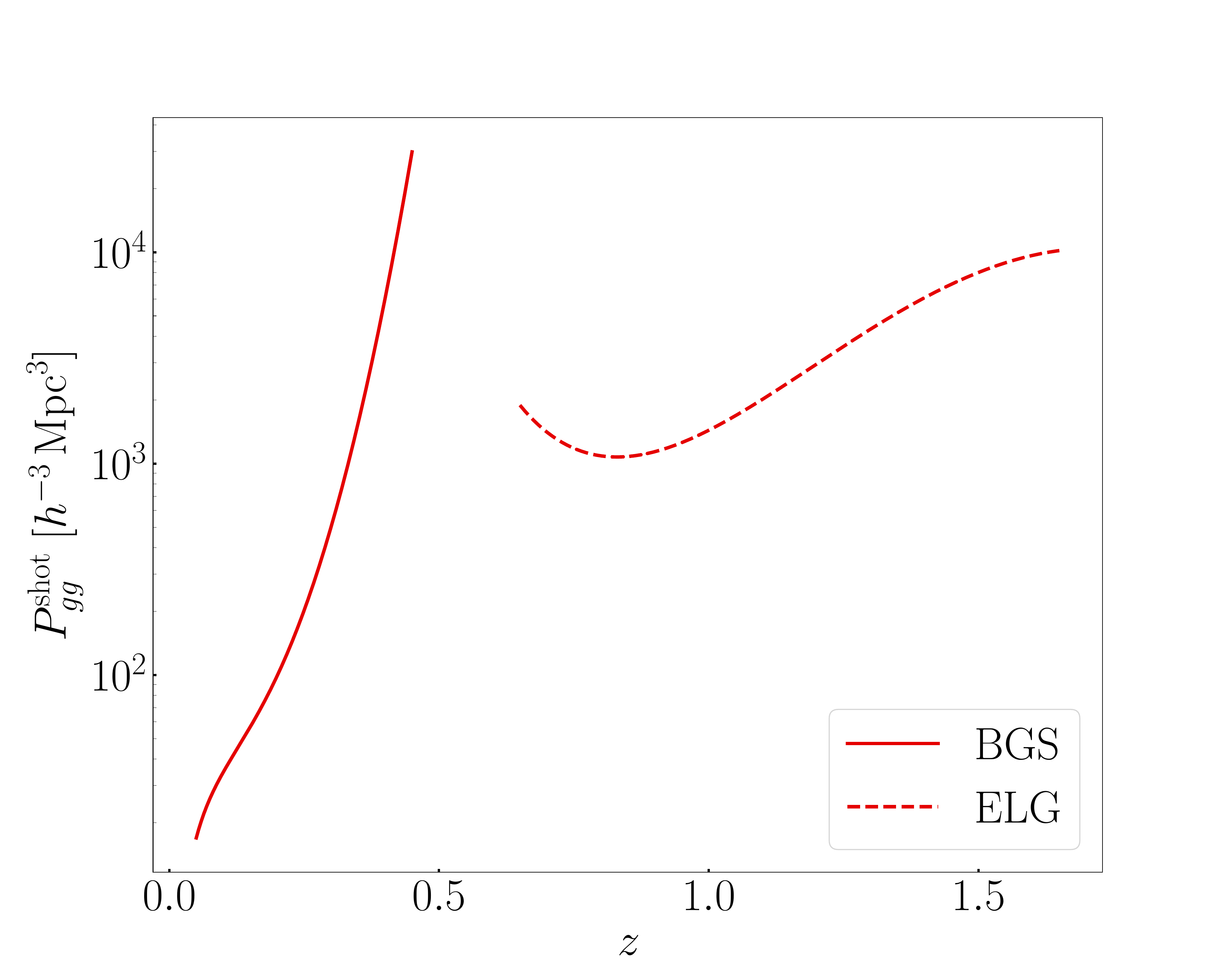} 
\includegraphics[width=5.9cm]{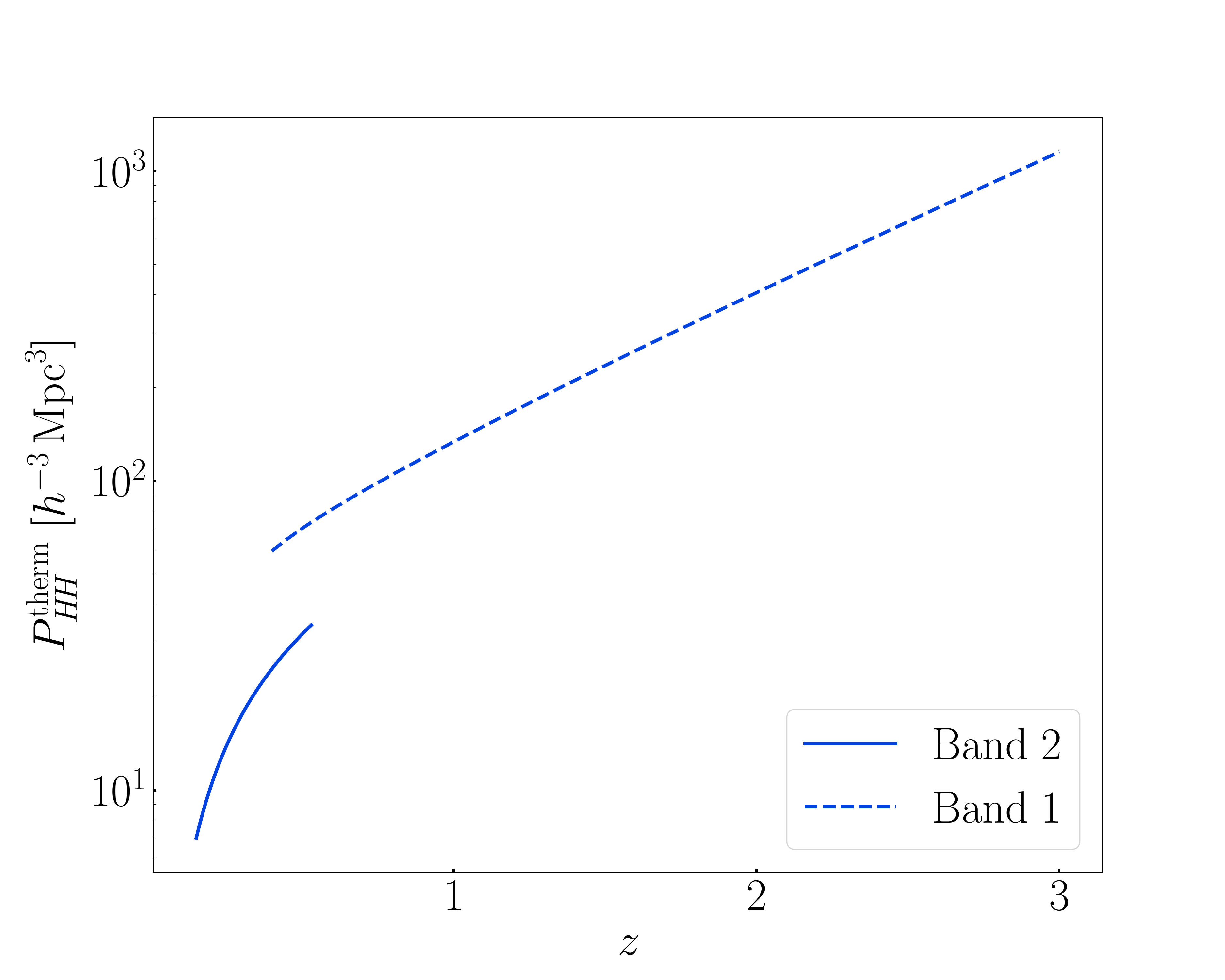} 
\caption{ Shot noise for galaxy surveys (\emph{left}) and thermal noise for IM surveys (\emph{right}).} \label{figNoise}
\end{figure}

The thermal noise in HI IM  depends on the sky temperature in the radio band, the survey specifications and the array configuration (single-dish or interferometer). For the single-dish  mode of SKAO-like surveys, the thermal noise power spectrum is  \cite{Bull:2014rha,Alonso:2017dgh,Jolicoeur:2020eup}:
\begin{equation}
P_{H\!H}^{\mathrm{therm}}(z) = \frac{\Omega_{\mathrm{sky}}}{2\nu_{21} t_{\mathrm{tot}}}\,\frac{(1+z) r(z)^{2} }{\mathcal{H}(z) }\,\left[\frac{T_{\mathrm{sys}}(z)}{\bar{T}_{H}(z)}\right]^2\, \frac{1}{N_{\rm d}}\,, 
\label{E1.32}
\end{equation}
where $\nu_{21}=1420\,\mathrm{MHz}$ is the rest-frame frequency of the 21\,cm emission,  $t_\mathrm{tot}$ is the total observing time, and the number of dishes is $N_{\mathrm{d}}=197$ (with dish diameter  $D_{\mathrm{d}}=15\,$m).
The system temperature is modelled as \citep{Ansari:2018ury}:
\begin{equation}
T_{\mathrm{sys}}(z) = T_{\rm d}(z)+T_{\rm sky}(z) =T_{\rm d}(z) + 2.7 + 25\bigg[\frac{400\,\mathrm{MHz}}{\nu_{21}} (1+z)\bigg]^{2.75} ~ \mathrm{K}, \label{E1.33}
\end{equation} 
where $T_{\rm d}$ is the dish receiver temperature (see \cite{Viljoen:2020efi}). The total signal is then 
\begin{equation}
\tilde{P}_{H\!H}(z,k,\mu) = P_{H\!H}(z,k,\mu) + P^{\rm shot}_{H\!H}(z) + P_{H\!H}^{\mathrm{therm}}(z)\approx P_{H\!H}(z,k,\mu) + P_{H\!H}^{\mathrm{therm}}(z)\,.\label{EEF1.34}
\end{equation}
The noise for all surveys is shown in \autoref{figNoise}.

In the case of the cross-power spectrum $P_{gH}$,   cross-shot noise  arises if the halos that host galaxies and HI overlap. Following the arguments given in \cite{Viljoen:2020efi,Casas:2022vik}, we assume that the cross-shot noise may be neglected. The total cross-power signal is then
\begin{equation}
\tilde P_{gH}=P_{gH}~~\mbox{with}~~ P_{gH}^{\rm shot}\approx 0\,. \label{cshot}
\end{equation}

\subsection{Intensity mapping beam and foregrounds}

HI IM surveys like the SKAO surveys have poor angular resolution, which results in power loss on small transverse scales, i.e. for large $k_\perp=(1-\mu^2)^{1/2}k$. This effect is typically modelled by a Gaussian beam factor \citep{Bull:2014rha}:
\begin{equation}
{\cal D}_{\rm b}(z,k,\mu)=\exp\left[-{(1-\mu^2)k^2 r(z)^2\theta_{\rm b}(z)^2\over 16\ln 2} \right] 
\quad\mbox{with}\quad \theta_{\rm b}(z) = 1.22\,{\lambda_{21}(1+z)\over D_{\rm d}}\,.
\label{e1.30}
\end{equation}

\begin{figure}[! ht]
\centering
\includegraphics[width=5.9cm]{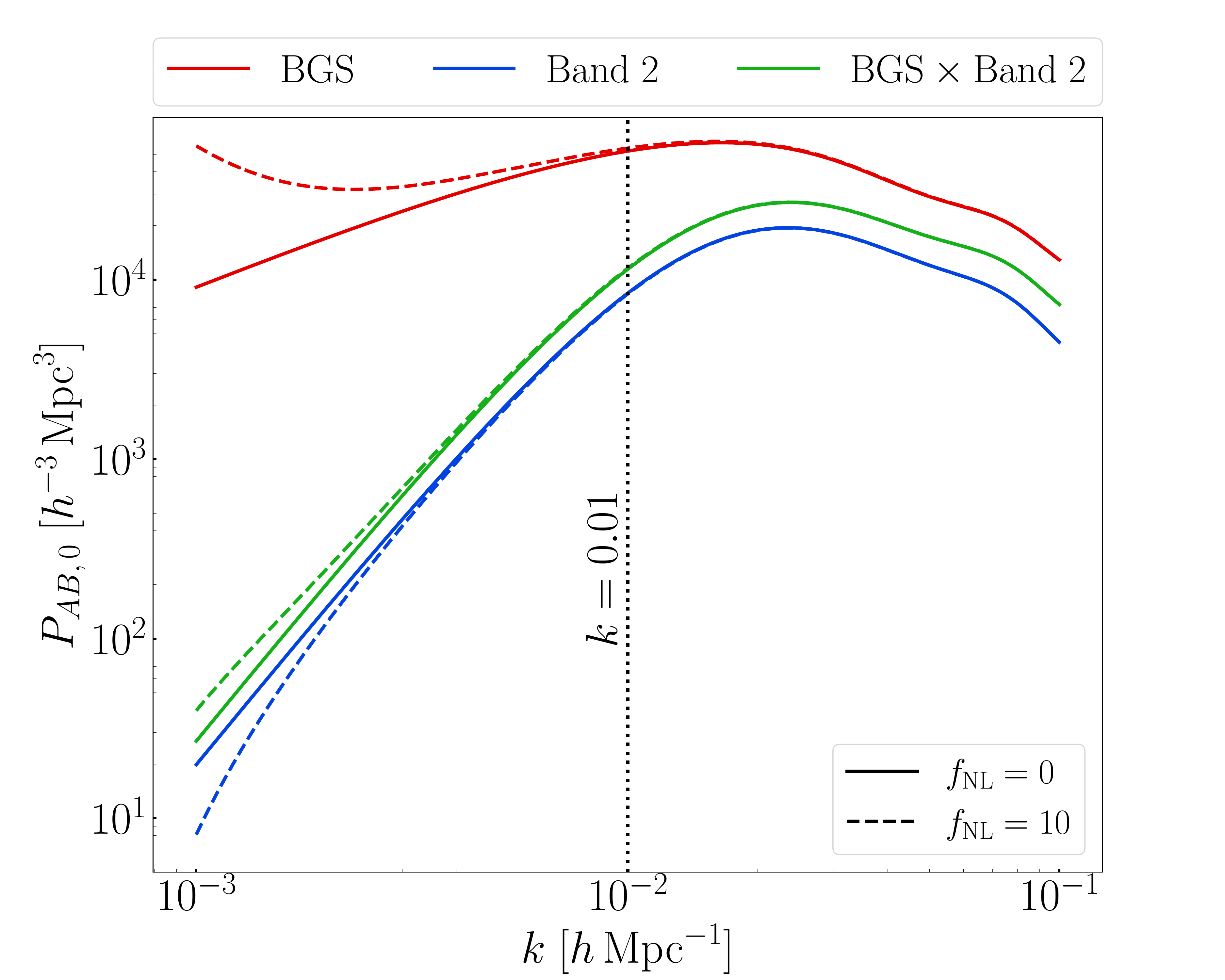} 
\includegraphics[width=5.9cm]{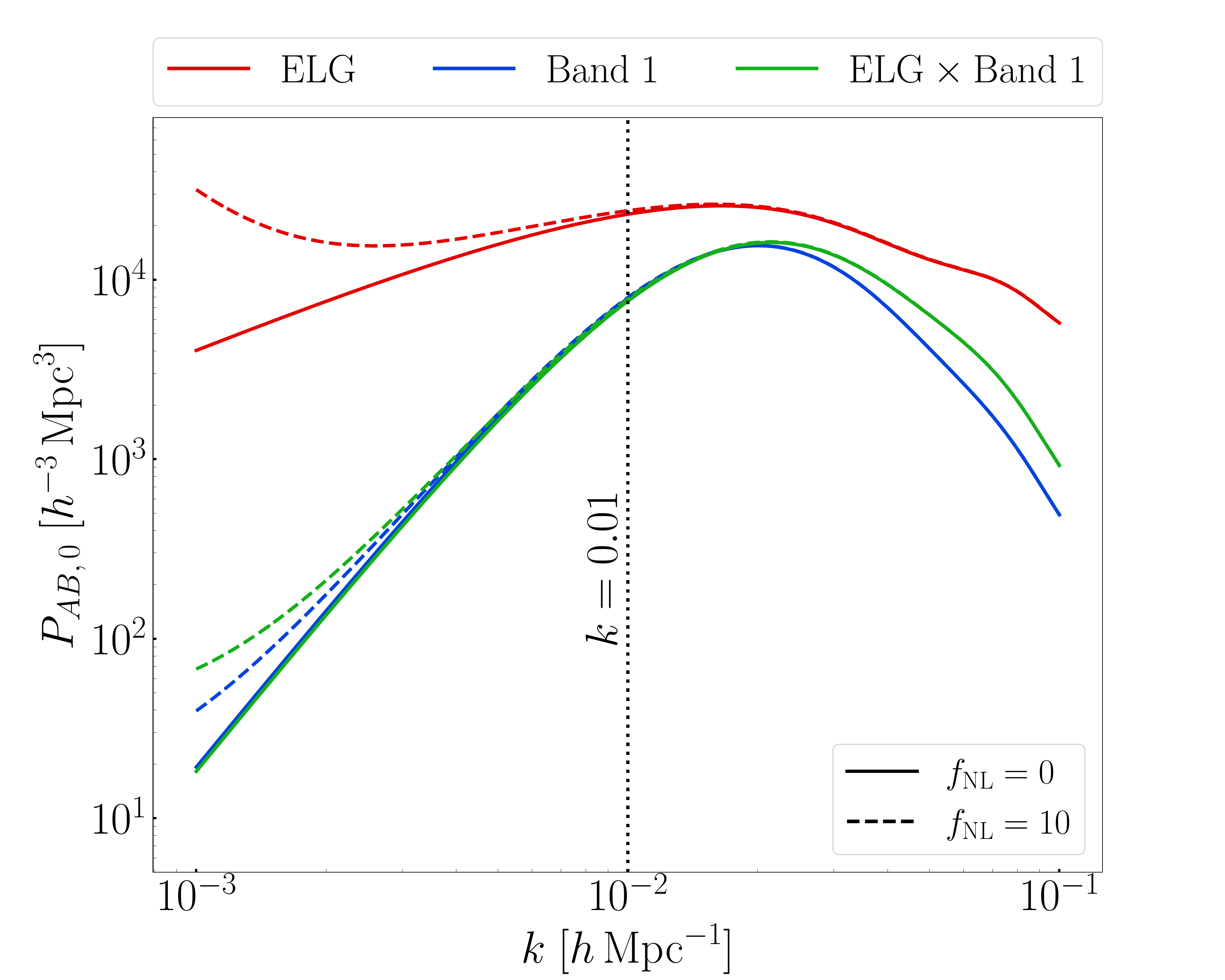} 
\caption{The monopole power spectra at $z=0.3$ (\emph{left}) and $z=1.0$ (\emph{right}). } \label{figMultipole}
\end{figure}

HI IM surveys are also contaminated by foregrounds  that are much larger than the HI signal. 
Since these foregrounds are spectrally smooth, they can be separated from the non-smooth signal on small to medium scales. However, on very large radial scales, i.e. for small $k_\|=\mu k$, the signal becomes smoother and therefore the separation fails.
A comprehensive treatment includes simulations of foreground cleaning of the HI signal (e.g. \cite{Cunnington:2020wdu,Spinelli:2021emp}). 
For a simplified Fisher forecast we can instead use a foreground avoidance approach, by excising the regions of Fourier space where the foregrounds are significant. 
 This means removing large radial scales, which can be modelled  via an exponential suppression factor \citep{Bernal:2019jdo,Cunnington:2022ryj}: 
\begin{equation}
{\cal D}_{\rm fg}(k,\mu) = {1-\exp\bigg[-\Big(\frac{\mu k}{k_{\rm fg}} \Big)^2 \bigg]\,.} \label{e1.311}
\end{equation}
We choose a typically used value: 
\begin{equation}
k_{\rm fg} = 0.01\,h \,{\rm Mpc}^{-1}
\,.  \label{efg} 
\end{equation}
\autoref{figMultipole}  illustrates the consequences of foreground avoidance for the monopoles of the HI power spectra, 
\begin{equation}
P_{AB,0}\,(z,k) = \frac{1}{2}\int_{-1}^{+1}\mathrm{d}\mu\,P_{AB}\,(z,k,\mu)\;.\label{monopole}
\end{equation}
Note the consequence of foreground avoidance for the monopole of the HI power spectra $P_{AH,0}$, which are suppressed on large scales.
It is clear that large-scale (small $k$) power in $P_{AH,0}$ is lost and that the effect of local PNG on these large scales is suppressed by foreground avoidance. Note that $f_{\rm NL}$ {\em reduces} the HI power on very large scales at $z=0.3$, since $b_H<1.25$, which leads to $b_{H\phi}<0$ in \eqref{e1.19}. Also note that the beam effect on $P_{AH,0}$, which tends to suppress small scale (large $k$) modes, is negligible at the low redshift, but becomes apparent at higher redshift.  \\

In summary, the effects of the radio telescope beam and radio foreground avoidance lead to the following modifications of the power spectra $P_{AH}$: 
\begin{eqnarray}
P_{H\!H}(z,{k}, \mu) &\rightarrow & {\cal D}_{\rm fg}(k,\mu)\,{{\cal D}_{\rm b}(z,k,\mu)}^{2}\,P_{H\!H}(z,{k}, \mu)\,, \label{EE1.31} \\
P_{gH}(z,{k}, \mu) &\rightarrow & {\cal D}_{\rm fg}(k,\mu)\,{\cal D}_{\rm b}(z,k,\mu)\,P_{gH}(z,{k}, \mu)\,. \label{EEQ1.31}
\end{eqnarray}
{We note that these equations are consistent with the recent paper \cite{Cunnington:2023jpq} on foreground cleaning via a transfer function, which shows that the effect of foregrounds is the same in the HI auto-power as in the cross-correlation power spectra with galaxies. However, we should also point out that the modelling of foreground removal is an ongoing project which has not yet reached maturity.}

\newpage
\section{Fisher forecast}\label{secfish}
For a multi-tracer combination of two dark matter tracers  $g$ and $H$, the data vector of  power spectra is
\begin{equation}
\bm{P} = \big( P_{gg}\,,\,  P_{gH}\,,\,  P_{H\!H}\big)\,.\label{e4.1}
\end{equation}
We exclude the noise from the data vector, since the noise is independent of the cosmological and nuisance parameters. 
(The noise appears in the  covariance, as given below.)
The standard cosmological parameters are measured on medium to small scales and are effectively independent of the ultra-large scale parameter $f_{\mathrm{NL}}$. 
Fixing their values could bias the best-fit value of $f_{\mathrm{NL}}$ but is unlikely to have any significant impact on $\sigma(f_{\mathrm{NL}})$.
We include in the Fisher analysis the cosmological parameters that directly affect the large-scale amplitude ($\sigma_{8,0}$) and shape ($n_s$) of the power spectrum.  We therefore consider the following cosmological and nuisance parameters:
\begin{equation}
\vartheta_{\alpha} = \big(\sigma_{8,0},\,n_{s},\,f_{\mathrm{NL}};\,b_{g0},\,b_{H0}\big)\,.\label{e4.2}
\end{equation}
Here we assume that the degeneracies between $b_A$ and $\sigma_{8,0}$, and between $\bar T_H$ and $\sigma_{8,0}$, have been broken by other surveys that focus on medium to small scales.

The covariance for the multi-tracer power spectra is given by \cite{Barreira:2020ekm,Karagiannis:2023}: 
\begin{equation} \label{e4.3}
\mathrm{Cov}(\bm{P},\bm P) =  \frac{k_{\rm f}^3}{2\pi k^{2} \Delta k}\,\frac{2}{\Delta \mu}\,
\begin{pmatrix}
\tilde P_{gg}^2 & & \tilde P_{gg}\tilde P_{gH} & & \tilde P_{gH}^2 \\ 
& & & & \\
\tilde P_{gg}\tilde P_{gH} & & \frac{1}{2}\big[\tilde P_{gg}\tilde P_{H\!H}+ \tilde P_{gH}^2 \big] & & \tilde P_{H\!H}\tilde P_{gH} \\
& & & & \\
\tilde P_{gH}^2 & & \tilde P_{H\!H}\tilde P_{gH} & & \tilde P_{H\!H}^2 
\end{pmatrix}\,.
\end{equation}
Here $\Delta k$ and $\Delta \mu$ are the  bin-widths for $k$ and $\mu$ and $k_{\mathrm{f}}$ is the fundamental mode, corresponding to the longest wavelength, which is determined  by the comoving survey volume of the redshift bin centred at $z$:
\begin{equation}
V(z) = \frac{\Omega_{\mathrm{sky}}}{3} \bigg[r\Big(z+\frac{\Delta z}{2}\Big)^{3} - r\Big(z-\frac{\Delta z}{2}\Big)^{3} \bigg]=
\bigg[\frac{2\pi}{k_{\mathrm{f}}(z)} \bigg]^3
\,. \label{e4.4}
\end{equation}
Then the multi-tracer Fisher matrix in a redshift bin is 
\begin{equation}
F_{\alpha \beta}^{\bm P} = \sum_{\mu=-1}^{+1}\,\sum_{k=k_{\mathrm{min}}}^{k_{\mathrm{max}}}\,\partial_{\alpha}\,\bm{P} \cdot \mathrm{Cov}(\bm{P},\bm P)^{-1} \cdot \partial_{\beta}\,\bm{P}^{\mathrm{T}}\,, \label{e4.6}
\end{equation}
where $\partial_{\alpha} = \partial\,/\,\partial \vartheta_{\alpha}$. We choose the bin-widths and $k_{\mathrm{min}}$ following \citep{Karagiannis:2018jdt,Yankelevich:2018uaz,Maartens:2019yhx}:
\begin{equation}
\Delta \mu = 0.04\,, ~~ \Delta k = k_{\mathrm{f}}\,, ~~ k_{\mathrm{min}} = k_{\mathrm{f}}\,.
\end{equation}
The smallest scale (largest $k$) is chosen  to ensure that linear perturbations remain accurate, since we do not require information from nonlinear scales {\cite{Smith:2002dz}}:
\begin{equation}
    k_{\mathrm{max}} = 0.08(1+z)^{2/(2+n_{s})}\,h\,\text{Mpc}^{-1}\,. \label{e4.7}
\end{equation}

\begin{figure}[! h]
\centering
\includegraphics[width=5.8cm]{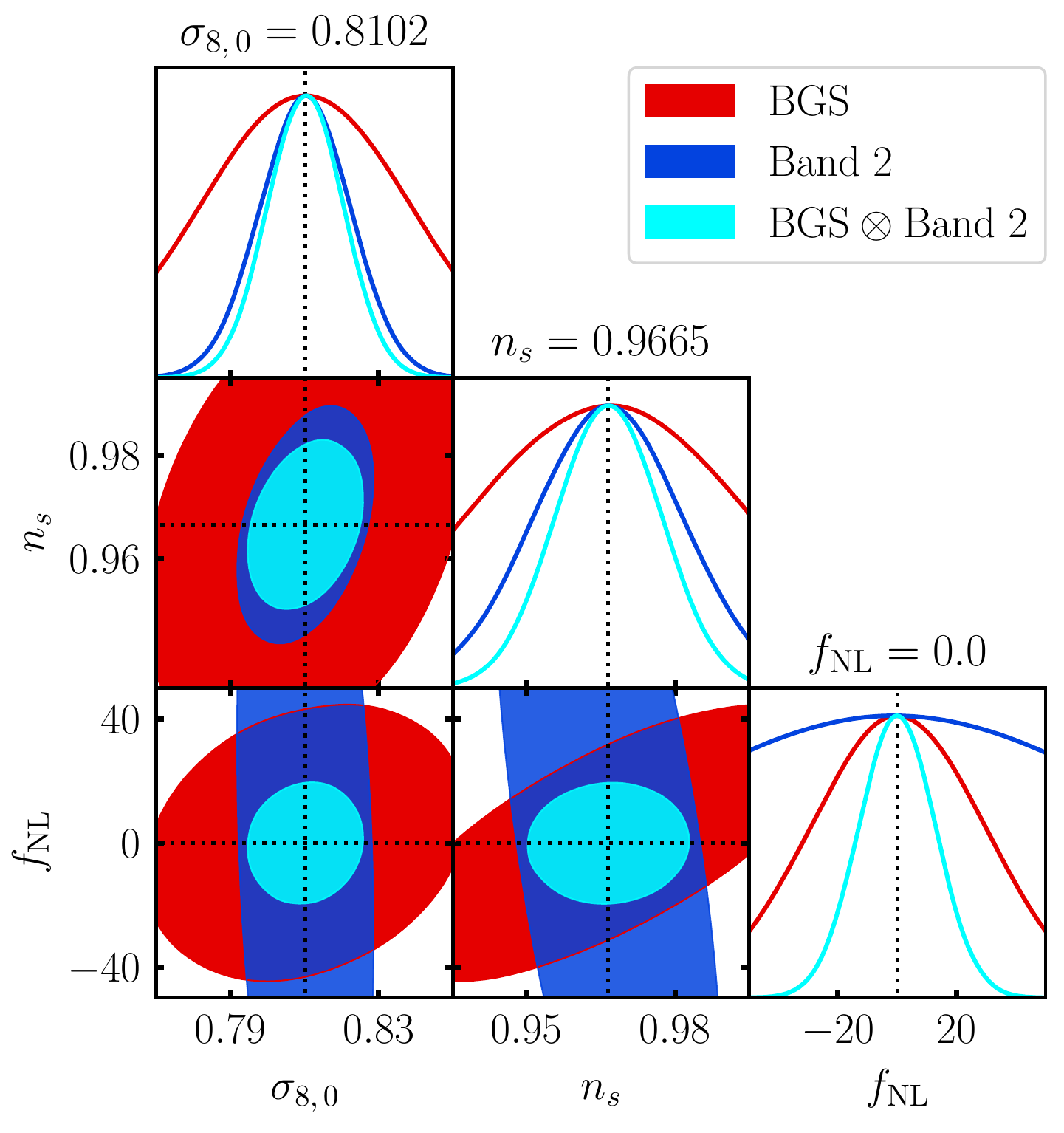} 
\includegraphics[width=5.9cm]{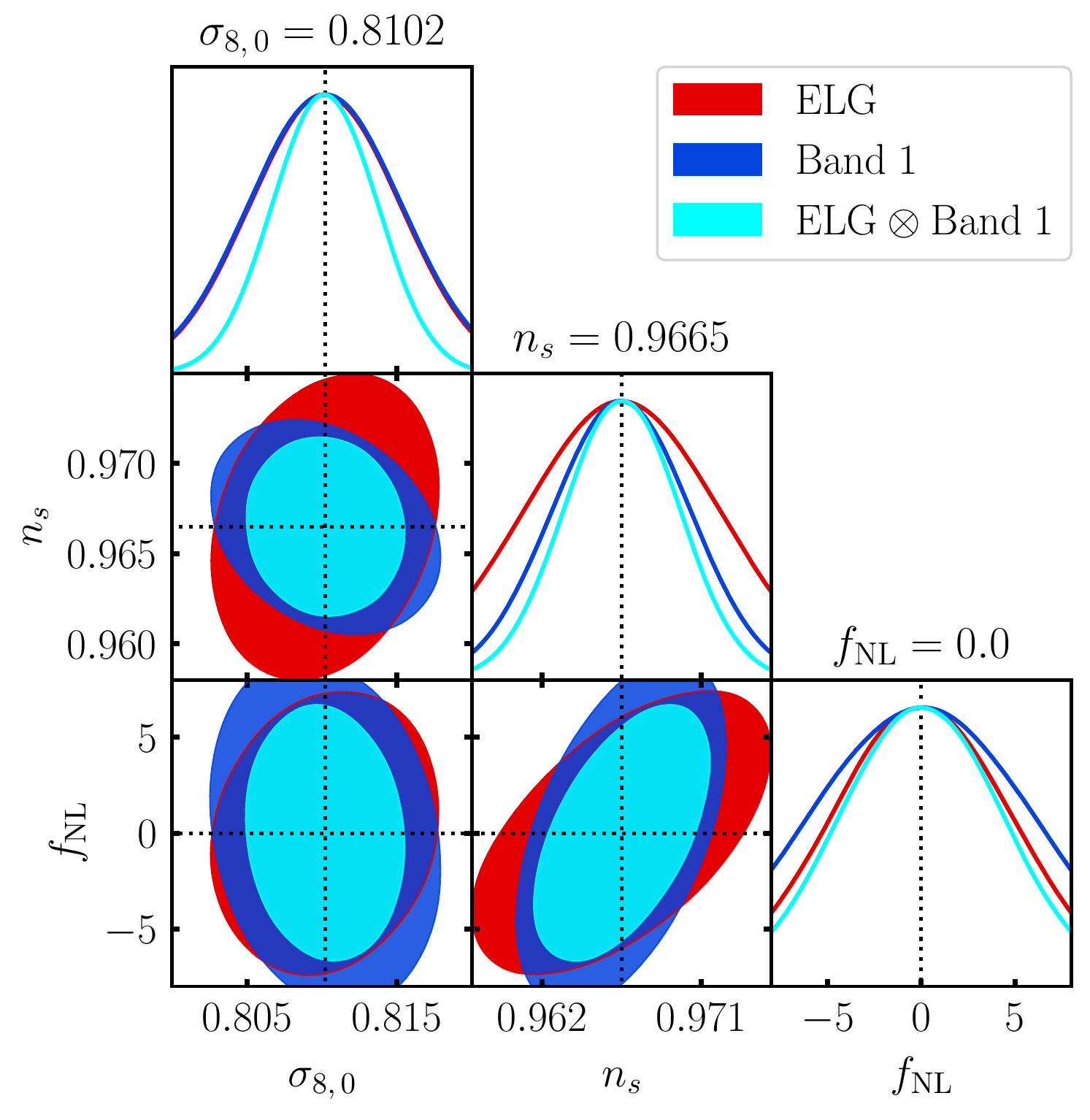} 
\caption{1$\sigma$ contours, including foreground avoidance in HI intensity mapping and marginalising over bias parameters.} \label{fig1}
\end{figure}

The galaxy and HI IM surveys do not have the same sky area and redshift ranges (see \autoref{tabInfo}). The overlap redshift ranges are obvious; for the sky areas,  we assume a nominal overlap area of 10,000\,deg$^2$. 
The multi-tracer applies in the overlap redshift range and overlap sky area. However,
we can add the independent Fisher matrices obtained in the non-overlapping regions of the two individual surveys to the multi-tracer Fisher matrix in the overlapping region \citep{Viljoen:2020efi}. In detail, the non-overlap region is  
\begin{itemize}
\item   the non-overlap  sky area of each survey, across the full redshift range of each survey;
\item the overlap sky area, across the non-overlap parts of the redshift ranges of each survey.  
\end{itemize}
Then the full  Fisher matrix (denoted by $g\otimes H$) is
\begin{equation}\label{fullf}
F_{\alpha \beta}^{g \otimes H} = 
F_{\alpha \beta}^{\bm P}(\text{overlap})+
F_{\alpha \beta}^{g}(\text{non-overlap}) + F_{\alpha \beta}^{H}(\text{non-overlap}) .  
\end{equation}
Figure~\ref{fig1} shows the 1$\sigma$ error contours computed from this Fisher matrix after marginalising over the bias parameters in \eqref{e4.2}. We use the fiducial values $\sigma_{8,0}=0.8102$, $n_{s}=0.9665$ and $f_{\mathrm{NL}}=0$, keeping other cosmological parameters fixed to their Planck 2018 best-fit values \citep{Planck:2018vyg}. Table~\ref{tab7} lists the marginalised $\sigma(f_{\mathrm{NL}})$ constraints, including the improvements (in parentheses) that follow when we ignore the HI IM foregrounds, i.e. when $k_{\rm fg}=0$. 
\newpage
{The results for BGS, Band 1 and Band 2 are consistent with \cite{Viljoen:2021ocx}, which uses the angular power spectrum rather than the Fourier power spectrum used here.
}
    
\begin{table}[!ht] 
\centering 
\caption{Cumulative marginalised error $\sigma(f_{\mathrm{NL}})$ for: the single-tracer surveys; the low-$z$ and high-$z$ multi-tracer pairs -- including single-tracer information from the non-overlap volumes as in \eqref{fullf}; and  the sum of the multi-tracer pairs.\\
Values in parenthesis correspond to ignoring foregrounds, i.e., with $k_{\mathrm{fg}}=0$.} \label{tab7} 
\vspace*{0.2cm}
\begin{tabular}{|l|l|} 
\hline 
Survey & 
 {$\sigma(f_{\mathrm{NL}})$} \\ \hline\hline 
BGS    & 29.4  \\
ELG    & 4.9  \\  \hline 
Band 2 & 95.7 \,(66.9) \\ 
Band 1 & 6.1 ~~(1.9) \\ \hline 
BGS $\otimes$ Band 2  & 13.0 \,(3.2) \\ 
ELG $\otimes$ Band 1 & 4.4 ~~(1.8) \\ \hline 
BGS $\otimes$ Band 2 + ELG $\otimes$ Band 1 & 4.0 ~~(1.5) \\ \hline
\end{tabular}
\end{table}
\section{Conclusion}

We applied a simplified Fisher forecast analysis in order to gain insights into the improvements on $\sigma(f_{\mathrm{NL}})$ from a new multi-tracer combination of large-scale structure surveys that will be made possible by the upcoming HI intensity mapping surveys with the SKAO. For the spectroscopic galaxy samples, we used mock surveys based on DESI. 
This allowed for multi-tracer pairs at low and high redshifts, in order to see the level of improved precision from the higher volume at high redshifts.

We included all available Fisher information from these mock surveys,  using the multi-tracer Fisher matrix on the overlap volume (sky area and redshift range) and the single-tracer Fisher matrices on the non-overlap volumes, as in \eqref{fullf}. 
Our results for $\sigma(f_{\mathrm{NL}})$ are summarised in \autoref{tab7}, 
which gives the single-tracer results for the 4 samples,  the full multi-tracer results from \eqref{fullf} for the low- and high-redshift pairs, and finally the results from the Fisher sum of these pairs. The impact of foreground avoidance can be seen from the results (in parenthesis) when foregrounds are ignored, corresponding to $k_{\rm fg}=0$ in \eqref{e1.311}.
 
\autoref{fig1} shows the $1\sigma$ contours
for the 3 cosmological parameters ($\sigma_{8,\,0}$, $n_{s}$, $f_{\mathrm{NL}}$), marginalising over the 2 bias nuisance parameters ($b_{g0}$, $b_{H0}$). 

As expected, the foreground avoidance severely degrades the constraining power of the HI intensity surveys. Without foregrounds, the high-redshift Band~1 survey would perform as well as the multi-tracer combination with the ELG survey: $\sigma(f_{\mathrm{NL}})=1.8$. 
The reduced precision on $f_{\rm NL}$ for single-tracer HI intensity constraints is consistent with the findings of \cite{Cunnington:2020wdu}, where foreground cleaning is simulated. However, using 
the multi-tracer with a spectroscopic galaxy survey helps to reduce the effect of foregrounds on $\sigma(f_{\mathrm{NL}})$, as shown by \autoref{tab7}.

The multi-tracer analysis is also expected to mitigate other systematics in the galaxy and HI intensity samples. For example, HI intensity mapping is also affected by radio frequency interference, receiver noise, calibration errors and polarisation leakage.
Incorporating these and the galaxy  systematics requires extensive simulations, beyond the scope of our paper.

\newpage \noindent
\textbf{Acknowledgements} \\
We thank Dionysis Karagiannis for very helpful comments. We are supported by the South African Radio Astronomy Observatory (SARAO) and the National Research Foundation (Grant No. 75415). 


\bibliography{reference_library}


\end{document}